%% file: main.tex
\documentclass[article]{IEEEtran}
\usepackage[intlimits]{amsmath}
\usepackage{amsfonts}
\usepackage{tikz}

\usepackage{cite}
\usepackage[normalem]{ulem}
\usepackage{circuitikz}
\usepackage{tkz-euclide}
\usetikzlibrary{angles}
\usetikzlibrary{positioning}
\usepackage{tikz-3dplot}
\usepackage[utf8]{inputenc}
\usepackage{xcolor}
\usepackage{graphicx}
\usepackage{lipsum}
\usepackage{soul}
\usepackage{algorithm}
\usepackage{algpseudocode}
\usepackage{setspace}
\usepackage{float}
\newfloat{algorithm}{t}{lop}

\usepackage{hyperref}
\hypersetup{
	colorlinks=true, % false: boxed links; true: colored links
	linkcolor=blue!70!black, % color of internal links
	citecolor=red!70!black, % color of links to bibliography
	filecolor=blue!70!black, % color of file links
	urlcolor=blue!70!black, % color of external links
}

\newcommand{\M}[1]{\mathbf{#1}}
\newcommand{\T}[1]{\mathrm{#1}}
\newcommand{\V}[1]{\boldsymbol{#1}}
\newcommand{\UV}[1]{\hat{\boldsymbol{#1}}}

\newcommand{\herm}{\T{H}}

\newcommand{\ie}{\textit{i}.\textit{e}.{}} 
\newcommand{\eg}{\textit{e}.\textit{g}.{}}

\newcommand{\TRs}[1]{}

\usetikzlibrary{shapes.symbols}

\title{Iterative Calculation of Characteristic Modes Using Arbitrary Full-wave Solvers}
\author{Johan Lundgren, \IEEEmembership{Member, IEEE}, Kurt Schab, \IEEEmembership{Member, IEEE}, Miloslav Capek, \IEEEmembership{Senior Member, IEEE},\\ Mats Gustafsson, \IEEEmembership{Senior Member, IEEE}, and Lukas Jelinek
\thanks{Manuscript received \today; revised \today. This work was supported by the Czech Science Foundation under project~\mbox{No.~21-19025M} and the Swedish Research Council (2017-04656).}% <-this % stops a space
\thanks{J. Lundgren and M. Gustafsson are with Lund University, Lund, Sweden, (e-mails: \{johan.lundgren, mats.gustafsson\}@eit.lth.se).}
\thanks{K. Schab is with Santa Clara University, Santa Clara, USA (e-mail: kschab@scu.edu).}
\thanks{M. Capek and L. Jelinek are with Czech Technical University in Prague, Czech Republic (e-mails: \{miloslav.capek,lukas.jelinek\}@fel.cvut.cz).}
}
\begin{document}
%\setcounter{page}{0}
%\onecolumn
%\includegraphics[height=\textheight,page=1]{Cover.pdf}	
%\twocolumn
%\newpage
\let\Algorithm\algorithm
\renewcommand\algorithm[1][]{\Algorithm[#1]\setstretch{1.4}}

\maketitle

\begin{abstract}
    An iterative algorithm is adopted to construct approximate representations of matrices describing the scattering properties of arbitrary objects. The method is based on the implicit evaluation of scattering responses from iteratively generated excitations. The method does not require explicit knowledge of any system matrices (\eg{}, stiffness or impedance matrices) and is well-suited for use with matrix-free and iterative full-wave solvers, such as FDTD, FEM, and MLFMA. The proposed method allows for significant speed-up compared to the direct construction of a full transition matrix or scattering dyadic. The method is applied to the characteristic mode decomposition of arbitrarily shaped obstacles of arbitrary material distribution. Examples demonstrating the speed-up and complexity of the algorithm are studied with several commercial software packages.%, and the implemented algorithm is freely available.
\end{abstract}

\begin{IEEEkeywords}
Antenna theory, eigenvalues and eigenfunctions, computational electromagnetics, characteristic modes, scattering.
\end{IEEEkeywords}

\section{Introduction}

\IEEEPARstart{C}{haracteristic} mode analysis centers on the decomposition of the scattering properties of arbitrary structures into a basis with convenient properties, see~\cite{apm-1-lau2022characteristic,apm-2-capek2021computational,apm-3-adams2022antenna,apm-4-li2022synthesis,apm-5-manteuffel2022characteristic} for a detailed review. While characteristic mode analysis is commonly associated with the study of perfectly conducting objects using the method of moments~\cite{HarringtonMautz_ComputationOfCharacteristicModesForConductingBodies}, characteristic modes of arbitrary structures can be calculated using any full-wave solver using scattering-based formulations~\cite{gustafsson2021unified_part1,gustafsson2021unified_part2,capek2022characteristic}. This approach involves only the forward solution of a series of scattering problems to construct matrices that describe the scattering properties of the system under study. %Several choices exist for defining incident and scattered fields, including spherical or plane waves~\cite{Kristensson_ScatteringBook}, meaning characteristic modes can be obtained using transition matrices or scattering dyadics. 

An object's transition matrix~\cite[\S 7.8]{Kristensson_ScatteringBook} or its scattering dyadic~\cite[\S 7.8.1]{Kristensson_ScatteringBook} can be constructed using a series of simulations carried out by any full-wave solver capable of determining scattered fields due to a known excitation. A simple approach to this procedure (discussed in~\cite{gustafsson2021unified_part1} and \cite{capek2022characteristic}), involves a number of full-wave solutions coinciding with the  selected dimension of the transition matrix or scattering dyadic representation. For large problems necessitating matrix-free methods (\eg{}, iterative FEM solvers, FMM, and FDTD), \cite[Ch. 11]{Jin_TheoryAndComputationOfElectromagneticFields}, the cost of performing each of these full-wave simulations is typically far higher than the cost of decomposing the final transition matrix or matrix representation of the scattering dyadic.

In this work, we consider an iterative algorithm for efficiently estimating the complete transition matrix or scattering dyadic of arbitrary scatterers. This method yields characteristic modes without explicit knowledge of any matrices used within a particular numerical method. The proposed technique represents a matrix-free method for computing characteristic modes in an arbitrary full-wave solver. Additionally, the algorithm readily allows for calculating all characteristic modes with modal significance above a predefined threshold, which may be favorable over algorithms producing a fixed number of modes in decreasing order of modal significance.

%\section{Scattering Operators $\M{T}$ and $\V{S}$}
\section{Transition Matrix and Scattering Dyadic}

Consider an obstacle~$\varOmega$ illuminated by an incident field~$\V{E}_\T{i}$ and producing a scattered field~$\V{E}_\T{s}$. The transformation between incident and scattered fields can be described by matrices by projecting both onto wave-specific bases. Utilizing regular (Bessel) and outgoing (Hankel) spherical vector waves to represent the incident and scattered field, the scattering problem is completely defined by the transition matrix~$\M{T}$~\cite{Kristensson_ScatteringBook}
\begin{equation}
    \M{T}\M{a} = \M{f},
    \label{eq:t-def}
\end{equation}
where $\M{a}$ and $\M{f}$ are collections of spherical vector wave expansion coefficients for the incident and scattered field, respectively. Alternatively, if the incident field is described by a spectrum of plane waves 
\begin{equation}
    \V{E}_\T{i}(\V{r}) = \int_{4\pi} \V{E}(\UV{r}') \T{e}^{-\T{j}k\UV{r}'\cdot\V{r}}\,\T{d}\Omega'
\end{equation}
and the scattered field is studied in terms of the far-field $\V{F}$ defined as
\begin{equation}
    \V{F} (\UV{r}) = \lim_{r\rightarrow\infty}\V{E}_\T{s}(\V{r})r\T{e}^{\T{j} k r},
    \label{eq:ff-def}
\end{equation}
then the scattering problem is completely described by the scattering dyadic $\V{S}$ as~\cite{Kristensson_ScatteringBook}
\begin{equation}
    \frac{4\pi\T{j}}{k}\int_{4\pi}\V{S}(\UV{r},\UV{r}')\cdot\V{E}(\UV{r}')\,\T{d}\Omega' = \V{F}(\UV{r}).
    \label{eq:s-def}
\end{equation}
Here $k$ is the background wave number, $\T{j}^2=-1$, $\UV{r}=\V{r}/r$, and $r=|\V{r}|$.  Applying a quadrature rule to approximate the above integral transforms the scattering dyadic description of the problem into matrix form, see~\cite{capek2022characteristic}. %Both of these operators, the transition matrix~$\M{T}$ and scattering dyadic~$\V{S}$, contain rich amounts of information regarding the object's characteristic quantities and are of principal importance in scattering-based characteristic mode analysis~\cite{gustafsson2021unified_part1,capek2022characteristic}.  

\section{Scattering-based Characteristic Mode Decomposition}

Characteristic modes are fields or currents excited by characteristic excitations which are solutions to an eigenvalue problem involving either the transition matrix or scattering dyadic~\cite{gustafsson2021unified_part1, capek2022characteristic}. In the case when spherical vector waves are used, characteristic excitations are given by expansion coefficient vectors $\M{a}_n$ satisfying~\cite{gustafsson2021unified_part1}
\begin{equation}
\M{T}\M{a}_n = t_n\M{a}_n,
\label{eq:eig-t}
\end{equation}
while in the case of a plane-wave basis, characteristic excitations are the plane-wave spectra~$\V{E}_n$ computed from~\cite{capek2022characteristic}
\begin{equation}
\int\limits_{4\pi} \V{S}(\UV{r},\UV{r}') \cdot \V{E}_n(\UV{r}') \, \T{d}\varOmega' =  t_n \V{E}_n(\UV{r}).
\label{eq:eig-s}
\end{equation}
Note that, by the definitions of the transition matrix~\eqref{eq:t-def} and scattering dyadic~\eqref{eq:s-def}, the eigenvectors of either of the above eigenvalue problems are simultaneously representations of both incident and scattered fields, \ie{}, both instances of $\M{a}_n$ may be replaced by $\M{f}_n$ in \eqref{eq:eig-t} and both instances of $\V{E}_n$ may be replaced by~$\V{F}_n$ in \eqref{eq:eig-s}. Additionally, the characteristic excitations~$\M{a}_n$ or~$\V{E}_n$ may be applied to reconstruct modal current densities and other modal quantities~\cite{gustafsson2021unified_part2, capek2022characteristic}.  The scattering-based characteristic mode problem in \eqref{eq:eig-s} is equivalent to impedance-based problem for lossless structures, with explicit algebraic links between the two approaches available for many common integral equation formulations~\cite{gustafsson2021unified_part1}.  Characteristic mode eigenvalues of impedance-based formulations, commonly denoted as $\lambda_n$~\cite{apm-2-capek2021computational}, are related to those of the scattering-based formulation via~\cite{gustafsson2021unified_part1}
\begin{equation}
t_n = -\frac{1}{1+\T{j}\lambda_n}.
\end{equation}

\section{Construction of Scattering Operators}

The transition matrix~$\M{T}$ and scattering dyadic~$\V{S}$ are not, in general, known in closed form for arbitrary scattering systems. However, construction of either quantity (or its discrete representation in the case of the scattering dyadic) can be carried out with full-wave simulations capable of mapping arbitrary incident fields onto the resulting scattered fields. The number of incident field configurations (\ie{}, spherical or plane waves) required to construct accurate representations of $\M{T}$ or $\V{S}$ is commonly dictated by the electrical size of an obstacle~\cite{gustafsson2021unified_part1, capek2022characteristic} and varies from tens to thousands. However, the resulting matrices are often of low rank and their explicit calculation could be circumvented by a matrix completion technique~\cite{DavenportRomberg_2016}, such as the one introduced in the following section, that constructs approximate representations of $\M{T}$ or $\V{S}$ with a low number of calls to a full-wave solver.

Throughout this letter, we schematically denote the process of calling a full-wave solver as $\mathcal{L}(\V{E}^\T{i}) = \V{E}^\T{s}$, where the input (excitation) field~$\V{E}^\T{i}$ and output (scattered) field $\V{E}^\T{s}$ are assumed to be expressed in bases appropriate for constructing the operator of interest. For example, a solver assembling an incident field out of spherical waves with weights contained in the vector $\M{a}$ and returning a scattered field expressed in terms of spherical wave weights $\M{f}$ may be written schematically as $\mathcal{L}(\M{a}) = \M{f}$
even if the underlying transition matrix~\eqref{eq:t-def} is not explicitly computed during the simulation process. A similar notation is adopted for constructing incident fields with plane-wave weights $\V{E}(\UV{r}')$ and calculating the resulting scattered far-fields $\V{F}(\UV{r})$.

\section{Iterative Algorithms}
\label{section:IterativeAlgo}
\begin{algorithm}
\caption{Estimation of the transition matrix}\label{alg:T}
\input{alg-1.tex}
\end{algorithm}

\begin{algorithm}
\caption{Estimation of the scattering dyadic}\label{alg:S}
\input{alg-2.tex}
\end{algorithm}

Procedures for estimating the transition or scattering dyadic matrix of an object using a set of full-wave evaluations are sketched in Algorithms~\ref{alg:T} and \ref{alg:S}, respectively. Both algorithms follow similar steps, though the implementation details differ in a few small, yet significant, ways. These differences are highlighted in the following description of the procedure, where steps of both algorithms are referenced in parallel.

In the setup steps~\ref{state:init-m} and \ref{state:init-ex}, an iteration index $m$ is initialized to zero while an excitation vector is initialized with random values. For the transition matrix algorithm, the vector $\M{a}_0$ is initialized with random complex numbers while in the scattering dyadic algorithm the excitation $\V{E}_0$ is a random complex function. %In practice, the scattering algorithm is implemented in a discretized setting~\cite{capek2022characteristic} such that this initialization is also carried out with a finite set of random complex numbers corresponding to complex weights on plane waves in predefined incident directions.

Step~\ref{state:while} represents a condition for terminating the iterating component of the algorithm. Options for this condition are discussed further at the end of this section. Within the iterating loop, excitations are normalized in step~\ref{state:norm} using either a vector norm (transition matrix algorithm) or an inner product defined over the unit sphere (scattering dyadic algorithm). The normalized excitation is then used to drive a full-wave simulation in step~\ref{state:fullwave}.%, where the incident and scattered fields are assumed to be appropriately translated to and from the bases required by the particular algorithm and the full-wave solver being used.

In step~\ref{state:update-t}, the transition matrix or scattering dyadic is estimated by a set of outer products between all previous excitations and scattered fields. This construction ensures that incident fields map to the observed scattered fields when the incident fields are orthogonal in an appropriate inner product. The estimate of the scattering dyadic or transition matrix is then used to generate an estimate of the set of eigenvalues $\{t_n\}_m$ in step~\ref{state:est-t} using the appropriate choice of eigenvalue problem in \eqref{eq:eig-t} or \eqref{eq:eig-s}.  

In steps~\ref{state:update-p} and~\ref{state:update-a}, the Gram-Schmidt procedure~\cite{GolubVanLoan_MatrixComputations} is used to generate a new excitation based on the previous scattered field with zero projection onto the set of previous excitations. The index $m$ is then incremented and the procedure repeats from step~\ref{state:norm} until one or more stopping criteria are met.  

\subsection{Implementation Considerations}
Note that the classical Gram-Schmidt procedure described in step~\ref{state:update-p} is unstable and the excitations lose orthogonality after a few iterations. The modified Gram-Schmidt procedure stabilizes the procedure~\cite{GolubVanLoan_MatrixComputations} and should be used in~\ref{state:update-p}. This modification also highlights the similarity with Arnoldi iteration~\cite{GolubVanLoan_MatrixComputations}, but is for conceptual simplicity not described in Algorithms~\ref{alg:T} and~\ref{alg:S}. Additionally, the algorithms may be re-initialized with a random excitation in the orthogonal complement to $\M{P}_m$ or $\V{P}_m$ to improve robustness. 

An example of a stopping criterion is detecting an updated excitation norm less than a predefined value. %, i.e.,
%\begin{equation}
%    |\M{a}_m|<\epsilon \quad\text{or}\quad\int |\V{E}_m(\UV{r})|^2\T{d}\Omega<\epsilon.
%\end{equation}
Alternatively, the routine may be terminated when all modes with modal significance above a certain threshold $t_\T{thr}$ are sufficiently converged, that is,
\begin{equation}
\underset{n}{\max} \left\{\frac{|\tilde{t}_{n,m} - \tilde{t}_{n,m-1}|}{|\tilde{t}_{n,m}|}\right\} < \epsilon,
\label{eq:error}
\end{equation}
where
\begin{equation}
\{\tilde{t}_{n,m}\} = \{t_{n,m}: |t_{n,m}|>t_\T{thr}\}
\end{equation}
and the selected set of eigenvalues $\{\tilde{t}_{n,m}\}$ is sorted in decreasing absolute value.
 
\subsection{Extension to Transient Simulations}
\label{sec:td}
The presented iterative algorithms consider transition matrix and scattering dyadic representations at a single frequency. However, in multi-frequency cases, such as FDTD simulations, where a wide frequency response follows each numerical evaluation, the algorithm could be altered to consider the entire frequency interval. Provided multi-frequency simulations in which all frequency points are excited with the same weighted incident field, steps~\ref{state:fullwave}--\ref{state:est-t} are evaluated for each frequency point. This information is compiled and processed to provide an updated incident field in steps \ref{state:update-p}--\ref{state:update-a}. One could consider specific points in the interval; the highest frequency point, as without any prior knowledge of the object, this is the region with the most potential modes; or possibly the frequency point in which the sum of the estimated modal significances is the highest. The latter option evaluated every iteration of the algorithm provided the best results.
%\begin{equation}
%    \argmax \limits_{f_p \in f} \sum |\{t_n\}_{m,f_p}|
%    \label{eq:nightmare}
%\end{equation}
%We found the latter to give the best results for our cases as in each iteration the frequency point of most modal importance is picked.

\section{Numerical Examples}
%In this section, we present how the iterative techniques could alleviate the numerical burden by drastically reducing the number of full-wave simulation evaluations. 
%The results will be shown for single frequency data based on MoM/FEM simulations as well as wide frequency FDTD simulations.
%In the examples, the full matrices already have been computed and are used to synthesize the structure response of arbitrary excitations.

\begin{figure*}
    % \centering
    % \begin{tikzpicture}[scale=0.9,transform shape]
    % \node at (0,0) {\includegraphics[clip,trim=0in 0in 0.5in 0in]{figures/Fig_FEKO_cylinder_1.pdf}};
    % \node at (6,0) {\includegraphics[clip,trim=0.2in 0in 0.5in 0in]{figures/Fig_FEKO_cylinder_2.pdf}};
    % \node at (12.4,0) {\includegraphics[clip,trim=0.2in 0in 0in 0in]{figures/Fig_FEKO_cylinder_3.pdf}};

    % \immediate\write18{convert figures/CylinderCSTPrint.png -transparent white Cylindertmp.png}
    % \begin{scope}[shift={(5.0,-0.8)},scale=.7]
    % \draw (0,0) node{\includegraphics[width= 3.4 cm]{Cylindertmp.png}};
    % \draw[<->] (-1.6,-0.79)--(0.88,-0.7) node[midway,below]{$42\,\mathrm{mm}$};
    % \end{scope}    % \end{tikzpicture}
    \includegraphics[width=\textwidth]{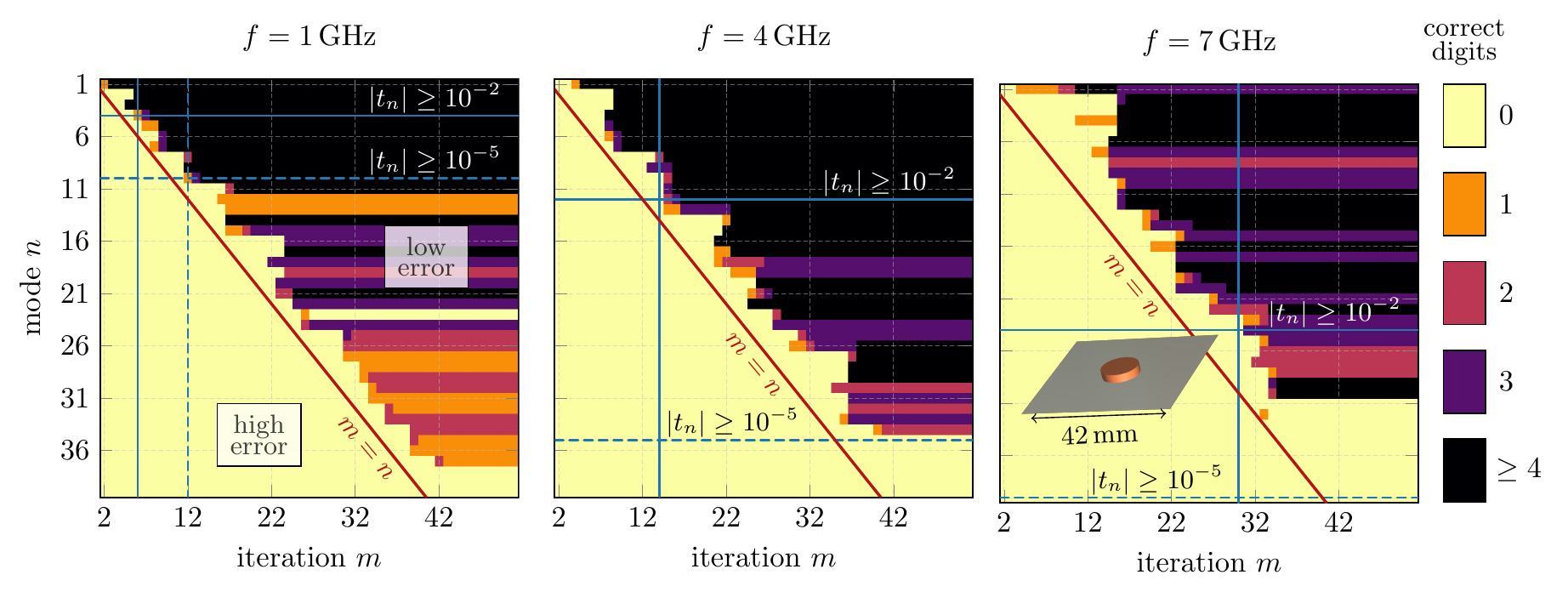}
    \caption{Number of correctly evaluated digits of eigenvalues $t_n$ depending on the number of iterations~$m$. The number of digits is evaluated from the comparison with the decomposition of fully constructed scattering dyadic as described in~\cite{capek2022characteristic}, $292$~plane waves being used. The full-wave solver used to gather the data is FEM-MoM hybrid implemented in Altair FEKO~\cite{feko}. The horizontal solid and dashed lines indicate characteristic modes reaching depicted thresholds of modal significance. The corresponding vertical solid and dashed line shows the number of iterations~$m$ required to acquire these modes.}
    \label{fig:single-freq-convergence}
\end{figure*}

As a demonstration of the proposed procedure, we consider the characteristic modes of a cylindrical dielectric resonator antenna on a finite ground plane~\cite{Huang2021}. The cylinder has a diameter of $10.5\,\mathrm{mm}$, a height of $2.3\,\mathrm{mm}$, and a relative permittivity of $\varepsilon_\mathrm{r}=38$. It is placed on top of the center of a perfectly-electrically-conducting square plate of side length $42\,\mathrm{mm}$, see Fig.~%\ref{fig:DielCylGeom}
\ref{fig:single-freq-convergence}

To calculate the characteristic modes of this structure, a realization of Algorithm~\ref{alg:S} was implemented using an iterative FEM-MoM hybrid solver within Altair FEKO~\cite{feko}. The discrete representation of the scattering dyadic is set up with Lebedev quadrature~\cite{capek2022characteristic} of a degree $146$ and a dimension of $N = 292$. Convergence of characteristic mode eigenvalues is demonstrated in Fig.~\ref{fig:single-freq-convergence}, where the number of correctly evaluated digits is plotted as a function of mode index and iteration number. Dark colors indicate convergence, while light colors correspond to modes that are not yet converged. For all three studied frequencies, the number of converged modes is approximately the number of iterations (see diagonal $m=n$ line), though an offset of approximately $10$ iterations is required to obtain any converged eigenvalues in the highest frequency case.  

Horizontal lines indicate the number of modes with modal significances above $10^{-2}$ and $10^{-5}$, while vertical lines indicate the number of iterations to resolve all modes above these thresholds. In practical applications of characteristic modes, only modes above the $|t_n| < 10^{-2}$ line are typically considered.  For the lowest frequency ($f=1~\T{GHz}$), the data in Fig.~\ref{fig:single-freq-convergence} show that only six simulations are required to accurately compute these dominant modes, much fewer than the $292$ solutions required to construct the entire matrix representing the scattering dyadic. This trend also holds for the two higher frequencies, where 14 and 30 iterations are required.

\begin{figure}
    \centering
    \includegraphics[width=0.45\textwidth]{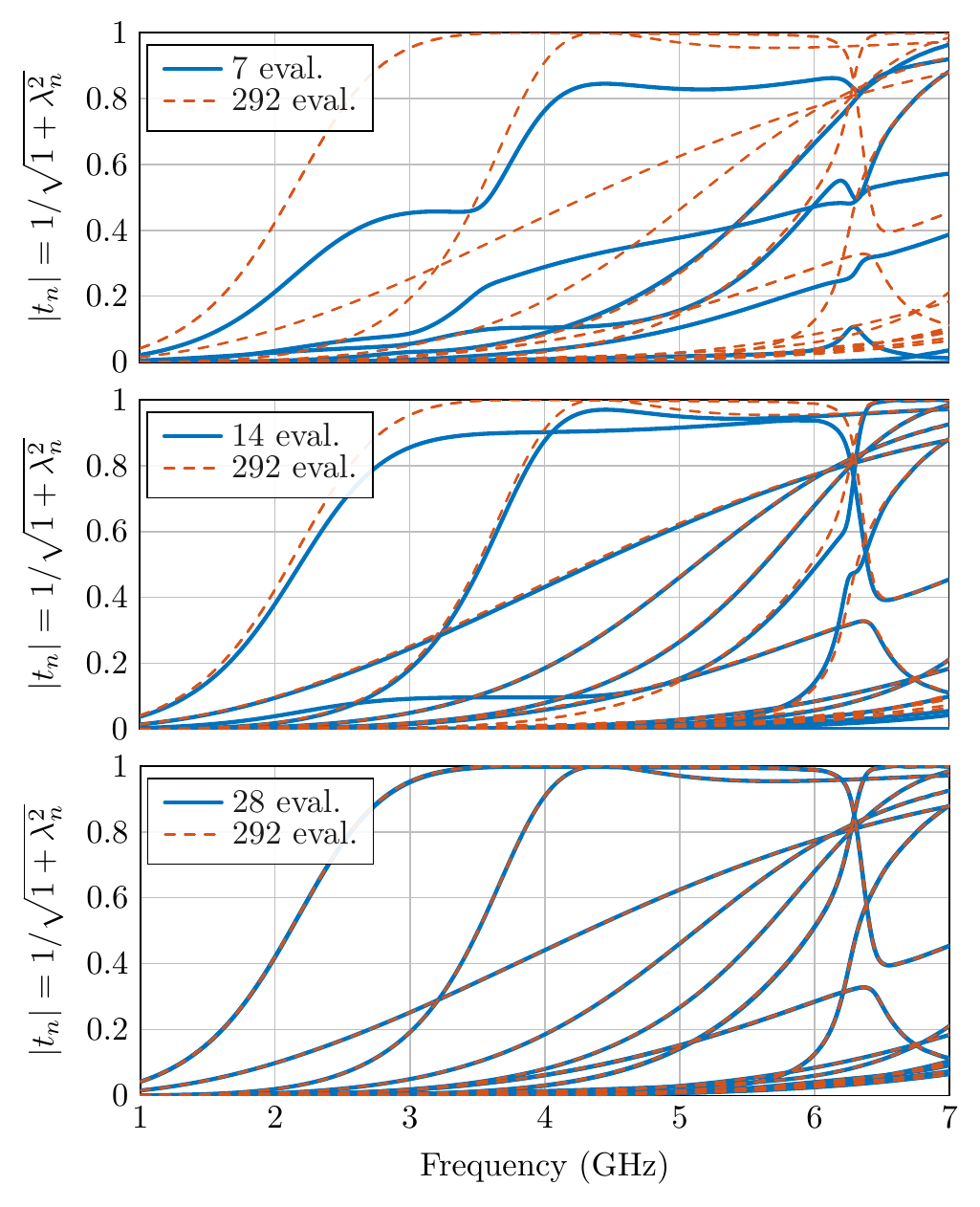}\\
    \caption{The 20 highest modal significances of a dielectric cylinder on a finite ground plane in the frequency interval $1$--$7\,\mathrm{GHz}$ obtained from a scattering matrix obtained using 292 excitations (dashed) and the estimates from $m \in \left\{7, 14, 28\right\}$ iterative evaluations (solid) with a discretized version of Algorithm~\ref{alg:S}.}
    \label{fig:DielCylGeom}
\end{figure}

Furthermore, a multi-frequency implementation of Algorithm~2 was constructed using the scattering matrix obtained through an FDTD solver, FIT in CST Studio Suite 2022~\cite{cst}. Similar to the FEM calculations above, the scattering matrix has dimension $N=292$ at each frequency point. 
The 20 highest modal significances from the complete scattering matrix are displayed in the frequency interval 1--7 GHz in the three panels of Fig.~\ref{fig:DielCylGeom} in dashed lines and serve as a visual reference for the results of the discretized Algorithm 2 extensions to transient simulations displayed in solid lines. Due to limitations in software the full matrices were used to generate the responses of simultaneous excitations with arbitrary complex weights. With 7 iterative excitations, top panel, the emergence of the modes can be seen with an only partial agreement. Using an intermediate of 14 excitations most significant modes are present but not fully reconstructed. Finally, in this numerical example, 28 excitations were sufficient to produce visually indistinguishable modes compared to the full-matrix, a reduction of $90\,\%$.

\section{Discussion}

To naïvely construct the transition matrix or a discrete representation of the scattering dyadic of dimension~$N$ requires $N$ full-wave evaluations. Estimates for the minimum dimension~$N_\T{min}$ required to resolve characteristic modes to a prescribed precision are available in the literature~\cite{gustafsson2021unified_part1, capek2022characteristic}, though these estimates tend to be quite conservative. A dominant trend, however, is that the dimension $N_\T{min}$ generally increases with the electrical size of the object under consideration. In the iterative approach presented here, the dominant $K$ eigenvalues of the scattering dyadic or transition matrix are estimated accurately using approximately $K_0 + K$ iterations, \ie{}, in linear time with some offset $K_0$. While the values of the parameters $N_\T{min}$ and $K_0$ are difficult to quantitatively predict due to their dependence on both geometric complexity and electrical size, it holds that if the scattering dyadic or transition matrix has dimension $N$ sufficient to resolve $K$ characteristic modes, then the iterative cost $K_0 + K$ is at worst equal to %\footnote{Equality here implies full reconstruction of the scattering dyadic or transition matrix with $N$ iterations.} 
$N$. This indicates that the iterative algorithm cannot perform worse, in terms of computational time, than full, explicit calculation. Nevertheless, the exact quantitative speed-up afforded by the iterative algorithm depends on the dimension $N$ of the matrices being approximated, which can be skewed by \emph{a priori} knowledge of the modal structure of the system being studied. For instance, incorrectly setting $N$ unnecessarily high will lead to artificially high speed-up, while prior knowledge of the number of significant modes might allow for the naïve approach to realize accurate results with small $N$, thus greatly reducing the observed speed-up afforded by the iterative approach.  

%In this way, it is most relevant to note that the net computational cost of the iterative algorithm depends only on a constant plus a linear scaling in the number of modes $K$ with modal significance above the stopping criteria threshold.  Having prior knowledge of this number may allow for the naïve approach to achieve comparable performance, but this knowledge is typically not available \emph{a priori} in practical characteristic mode analyses. Hence the speed-up afforded by the iterative approach can be significant in realistic calculation settings.

While based on a fundamentally different formulation of characteristic modes, the proposed iterative approach shares some motivational aspects with matrix-free methods for computing characteristic modes of large structures~\cite{dai2016large}, where matrix-vector multiplications are implemented using MLMFA within iterative eigenvalue algorithms to avoid the high cost of computing, storing, and inverting large impedance matrices.  In contrast to that work, however, the approach taken here utilizes an iterative approach to reduce the high computational burden of computing the scattering matrix, rather than circumventing its computation altogether.

%\bibliographystyle{IEEEtran}
%\bibliography{main}

\end{document}

%% file: alg-1.tex
\begin{algorithmic}[1]
\State\label{state:init-m} $m = 0$
\State\label{state:init-ex} $\M{a}_0 \gets \T{rand}$
\While{stopping criteria are not met}\label{state:while}
\State \label{state:norm}$\M{a}_m \gets \M{a}_m/|\M{a}_m|$
\State\label{state:fullwave} $\M{f}_m \gets \mathcal{L}(\M{a}_m)$
\State\label{state:update-t} $\M{T}_m \gets \sum_{p\leq m} \M{f}_p\M{a}_p^{\herm}$
\State\label{state:est-t} $\{t_n\}_m\gets \T{eig}(\M{T}_m)$
\State\label{state:update-p} $\M{P}_m \gets \sum_{p\leq m} 
\M{a}_p\M{a}_p^{\herm}$
\State\label{state:update-a} $\M{a}_{m+1} \gets \M{f}_m - \M{P}_m\M{f}_m$
\State $m \gets m+1$
\EndWhile
\end{algorithmic}

%% file: alg-2.tex
\begin{algorithmic}[1]
\State $m = 0$
\State $\V{E}_0(\UV{r}) \gets \T{rand}$
\While{stopping criteria are not met}
\State $\V{E}_m \gets \V{E}_m / \sqrt{\int |\V{E}_m(\UV{r})|^2 \, \T{d}\Omega}$
\State $\V{F}_m(\UV{r}) \gets \mathcal{L}(\V{E}_m(\UV{r}'))$
\State $\V{S}_m(\UV{r},\UV{r}') \gets \sum_{p\leq m} \V{F}_p(\UV{r})\V{E}^{*}_p(\UV{r}')$
\State $\{t_n\}_m\gets \T{eig}(\V{S}_m)$
\State $ \V{P}_m(\UV{r},\UV{r}') \gets \sum_{p\leq m} \V{E}_p(\UV{r})\V{E}^{*}_p(\UV{r}')$
\State $\V{E}_{m+1} \gets \V{F}_m - \int \V{P}_m(\UV{r},\UV{r}')\cdot\V{F}_m(\UV{r}') \, \T{d}\Omega'$
\State $m \gets m+1$
\EndWhile
\end{algorithmic}